\documentclass[12pt]{article}
\usepackage{amssymb}
\usepackage{graphicx}
\usepackage{psfrag}
\usepackage{amsfonts}
\usepackage{epsfig}
\newcommand{\sef}{\sin^2 \theta_{eff}^{lept}}
\newcommand{\ini}{\begin{equation}}
\newcommand{\fin}{\end{equation}}
\newcommand{\sms}{\hat{s}^2}

\newcommand{\es}{s_{eff}^2}
\newcommand{\ec}{c_{eff}^2}
\newcommand{\drwc}{\Delta\hat{r}_W}
\newcommand{\droc}{\Delta\hat{\rho}}
\newcommand{\dkc}{\Delta\hat{k}}
\newcommand{\dre}{\Delta r_{eff}}
\newcommand{\msbar}{\overline{MS}}
\newcommand{\almuh}{\hat{\alpha}(\mu)}
\newcommand{\sinmuh}{\sin^2\hat{\theta}_W(\mu)}
\newcommand{\bite}{\begin{itemize}}
\newcommand{\eite}{\end{itemize}}
\newcommand{\smu}{\hat{s}^2(\mu)}

\newcommand{\dah}{\Delta\alpha_h^{(5)}}

\begin{document}

\hyphenation{re-nor-ma-li-za-tion}

\begin{flushright}
NYU-TH/01/06/01\\
hep-ph/0106094
\end{flushright}

\vspace{0.5cm}
\begin{center}
{\Large \bf Novel Approach to Renormalize the Electroweak Sector of the 
 Standard Model}\\
\vspace{0.2cm}
{\large A.~Ferroglia\footnote{e-mail: andrea.ferroglia@physics.nyu.edu}, 
G.~Ossola\footnote{e-mail: giovanni.ossola@physics.nyu.edu}, 
and A.~Sirlin\footnote{e-mail: alberto.sirlin@nyu.edu}.}

\vspace{0.5cm}
{\it Department of Physics, New York University,\\
4 Washington Place, New York, NY 10003, USA.}
\end{center}
\bigskip

\begin{center}
\bf Abstract 
\end{center}
{ \small We discuss a novel approach to renormalize the Electroweak Sector of the Standard
Model, in which $\sef$, measured at the $Z^0$ peak, is identified with the basic 
renormalized electroweak mixing parameter. This approach shares the desirable 
convergence properties of the $\msbar$ scheme and provides a framework for calculations 
that are strictly independent of the electroweak scale in finite orders of
perturbation theory. We illustrate the method with updated precise calculations 
of $\sef$ and $M_W$, as functions of $M_H$, and comment on the implications 
of these calculations for the Higgs boson search and the theoretical 
prediction of $M_W$.}

\newpage

\begin{figure}[p]
\centering
\psfrag{m}{ {\small $\mu\ \left[ GeV \right]$}}
\psfrag{s}{{\footnotesize $\sef$ }}
\resizebox{12cm}{7cm}{\includegraphics{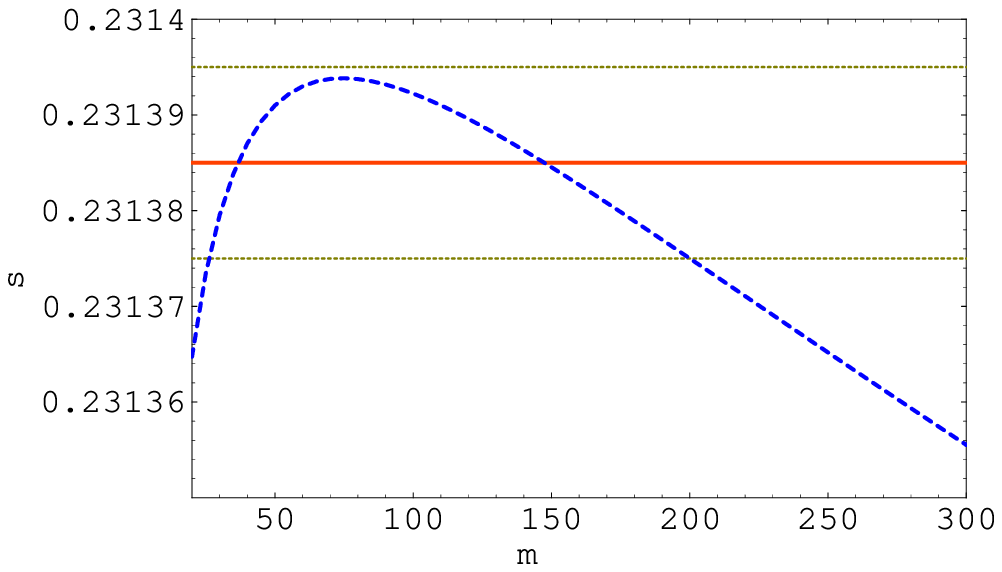}}
\caption{Scale dependence of $\es$ in the $\overline{MS}$ (dashed line)
and EFF (solid line) schemes for $M_H = 100 \ GeV$ 
and the input parameters listed in Table \ref{tab1}.
The light-dotted lines define a range of $\pm 1 \times 10^{-5}$ around the EFF result.}
\label{fig1}
\end{figure}

\begin{figure}
\centering
\psfrag{m}{ {\small $\mu\ \left[ GeV \right]$}}
\psfrag{w}{{\small $M_W$} {\small $\left[ GeV \right]$}}
\resizebox{12cm}{7cm}{\includegraphics{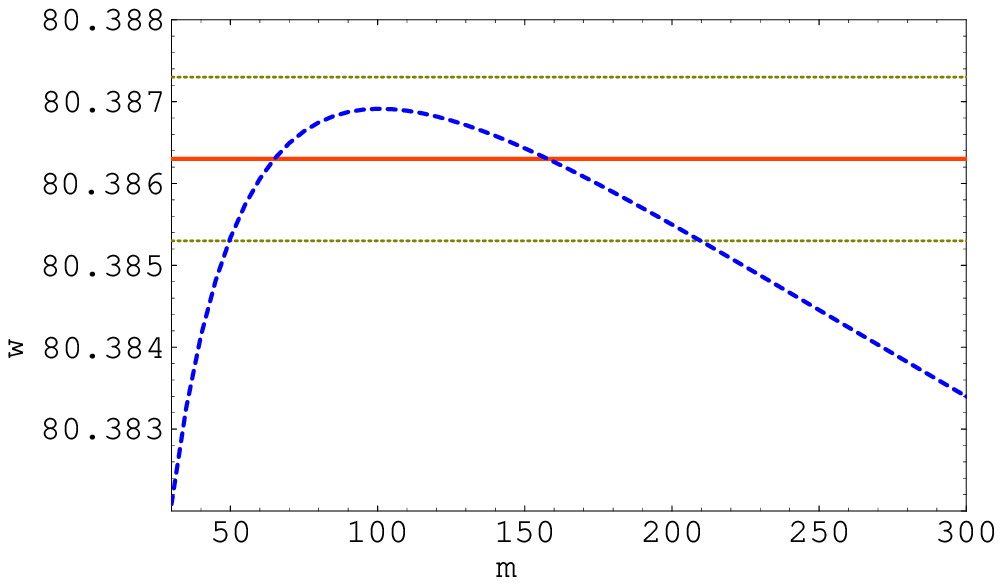}}
\caption{Scale dependence of $M_W$ in the $\overline{MS}$ (dashed line)
and EFF (solid line) schemes for $M_H = 100 \ GeV$ 
and the input parameters listed in Table \ref{tab1}. 
The light-dotted lines define a range of $\pm 1 \ MeV$ around the EFF result.}
\label{fig2}
\end{figure}

Current precise calculations in electroweak physics are based on a number of renormalization
schemes. The on-shell \cite{c1,c2} and $\msbar$ formulations \cite{c3,c4,c5,c6,c7}, which
are the most frequently employed, present a number of relative advantages and disadvantages.
The on-shell approach is very ``physical'' in the sense that it identifies the renormalized 
parameters with observable and, therefore, scale-independent quantities, such as $\alpha$,
$M_Z$, and $M_W$. The $\msbar$ calculations, on the other hand, have very desirable 
convergence properties. The reason is that they follow closely the structure of the 
unrenormalized theory and, in this way, avoid large finite corrections that frequently 
emerge in the renormalization process. They also involve parameters, such as $\almuh$ and 
$\sinmuh$, which are inherently scale dependent, and play a crucial role in the analysis
of grand unification.\\
On the other hand, since practical evaluations of important observables, such as $\sef$
and $M_W$, involve a truncation of the perturbative series, the use of the $\msbar$
scheme necessarily leads to a residual dependence on the electroweak scale $\mu$,
thus generating a source of ambiguity.\\
Very recently, a framework has been proposed that retains the desirable convergence 
properties of the $\msbar$ scheme and, at the same time, leads to calculations that are 
strictly independent of the electroweak scale in finite orders of perturbation theory \cite{c8}.
In the present communication, we review some of the highlights of this approach and illustrate
it with updated precise calculations of $\sef$ and $M_W$.
We also comment on the implications of these calculations for the upper bound for the Higgs 
boson mass $M_H$, and the theoretical prediction of $M_W$ based on the observed value of
$\sef$.\\
An important feature of the proposed scheme is that $\sef$, obtained from the precise 
measurements of the various asymmetries at the $Z^0$ peak, is identified with the 
basic renormalized electroweak mixing parameter. It is related to the accurately
known parameters $\alpha$, $G_\mu$, and $M_Z$ by
\ini \label{equ1}
\es \ec = \frac{A^2}{M_Z^2 \left( 1 - \dre \right)} \ ,
\fin
where $\es$ is an abbreviation for $\sef$, $A^2=\pi \alpha / \sqrt{2} G_\mu$ and $\dre$
is the relevant radiative correction. The one-loop approximation to $\dre$ has been recently
used to discuss the mass scale of new physics in the Higgs-less scenario \cite{knie} 
and the evidence for electroweak bosonic corrections in the SM \cite{tenyears}. \\
The main points in our strategy to evaluate $\dre$ beyond the one-loop approximation are:
\bite  
\item[i)]  Since current calculations of $\es$ incorporate two-loop effects enhanced 
by powers $(M_t^2/M_Z^2)^n$ (with $n=1,2$), we first express $\dre$ in terms of a 
set of corrections $\drwc$, $\droc$, $\dkc$ and $\hat{f}$, involving appropriate combinations 
of self-energies, vertex and box diagrams \cite{c3,c4,c5,c6,c7}, for which the irreducible 
contributions of this order have been already evaluated \cite{c9,c10,c11}.
\item[ii)] In order to ensure the absence of a residual electroweak-scale dependence, 
we use scale-independent couplings, such as $e^2$, $\es$, $G_\mu$, $M_Z^2$, 
retain only two-loop effects enhanced by factors  $(M_t^2/M_Z^2)^n$ ($n=1,2$),
since other two-loop corrections are not completely known, and employ a single 
definition of the electroweak mixing parameter, which we identify with $\es = 1- \ec$.
\eite  
In particular, the $\msbar$ parameter $\smu \equiv \sinmuh$ is expressed in terms of $\es$
by means of the relation \cite{c7}
\ini \label{equ2}
\es = \left[  1 + \frac{\hat{e}^2}{\sms} \dkc \left( M_Z^2, \mu \right)\right]\smu \ ,
\fin
where $\dkc(q^2, \mu)$ is the relevant electroweak form factor evaluated at
the momentum transfer $q^2=M_Z^2$, and $\mu$ is the 't-Hooft scale.\\
In this way, we obtain the relation \cite{c8}
\begin{eqnarray} \label{equ3}
\dre & = &\drwc - \frac{e^2}{\es} \left[\Delta \hat{\rho} -
\Delta \hat{k} \left( 1-\frac{\es}{\ec}\right)
\right]  \nonumber \\
& - & \frac{e^2}{\es} \ x_t \left[ 2 \Delta \hat{\rho} -
\left( \Delta \hat{\rho} \right)_{lead} - \hat{f} + \Delta \hat{k} 
\frac{\es}{\ec}\right] 
 \ ,
\end{eqnarray}
where $x_t = 3 G_\mu M_t^2/ (8 \sqrt{2} \pi^2)$ and $\left( \Delta \hat{\rho} \right)_{lead} 
= \left( 3 / 64 \pi^2 \right) M_t^2/ M_W^2 $ are the leading one-loop contributions to
 $ \left( \hat{e}^2/ \sms\right) \Delta \hat{\rho} $ and $\droc$, respectively, and we
have neglected two-loop contributions not enhanced by powers of  $(M_t^2/M_Z^2)^n$ ($n=1,2$).\\
The corrections $\drwc$, $\droc$, $\dkc$ and $\hat{f}$ depend also on $c^2=M_W^2/M_Z^2$, where 
$c^2$ is an abbreviation for the on-shell parameter $\cos^2{\theta_W}$ \cite{c1,c2}. In order
to ensure the exact cancellation of scale dependences among various contributions, it is
important to employ everywhere the same version of the electroweak mixing parameter which, 
as explained before, we identify with $\ec$. To achieve this, we proceed as follows:\\
a) $M_W^2$ is replaced everywhere by $c^2 M_Z^2$,
b) in all two-loop contributions we substitute $c^2 \to \ec$, since the difference is of third 
order,
c) in the one-loop terms we perform a Taylor expansion, exemplified by
\ini \label{equ4}
\Delta \hat{\rho} \left( \ec, c^2 \right) =   
\Delta \hat{\rho} \left( \ec, \ec \right) + \left.\frac{\partial \Delta \hat{\rho}}
{\partial c^2} \right|_{c^2 = \ec} \left( c^2 - \ec \right) +\cdots \ .
\fin
It suffices then to replace $c^2-\ec$ by its one-loop evaluation, expressed in terms
of scale-independent couplings.\\
The corresponding expression for the calculation of $M_W$ is given by \cite{c8}
\ini \label{equ5}
M_W^2/M_Z^2 = \ec \left\{ 1 - \frac{8 \ G_\mu}{\sqrt{2}}\ M_Z^2 \ c^2 \left[
\Delta \hat{\rho} + \frac{\es}{\ec} \Delta \hat{k}
\left( 1 -x_t\right) - \hat{f} x_t \right]\right\}^{-1} \ .
\fin
Again, $c^2$ is expressed in terms of $\ec$ in the manner explained above. An 
interesting feature of this formulation is that in Eqs.(\ref{equ1},\ref{equ2})
the calculation of $\es$ is carried out independently of $M_W$, while the 
$\es$ results are employed in Eq.(\ref{equ5}) to calculate $M_W$.
For brevity, we refer to this approach as the Effective Scheme of Renormalization (EFF).\\
In order to illustrate our results, we employ
$G_\mu = 1.16637  \times 10^{-5} \  GeV^{-2}$,
$M_Z = 91.1875 \ GeV$, $M_t = (174.3 \pm 5.1) \ GeV$, 
 $\hat{\alpha}_s \left( M_Z \right) = 0.118 \pm 0.002$, QCD corrections based on 
the $\mu_t$-parameterization \cite{c10,c12}, and two recent determinations of $\dah$:
$\dah = 0.02761 \pm 0.00036$  \cite{dah1} and $\dah = 0.02738 \pm 0.00020$ \cite{dah2}
(one of the ``theory-driven'' calculations).
In Table \ref{tab1} we present updated evaluations of $\es$ and $M_W$, based on the Effective 
Scheme, as functions of $M_H$, using the central values of the input parameters and  
$\dah = 0.02761$. In Table \ref{tab2}, we give the corresponding results for   
$\dah = 0.02738$. Figs. \ref{fig1} and \ref{fig2} compare the scale dependence of 
the $\msbar$ calculations of $\es$ and $M_W$ \cite{c10,c11} with the scale-independent Effective 
approach, for $M_H = 100 \ GeV$ and $\dah = 0.02761$.
We note that for the value $\mu=M_Z$ selected in Refs.\cite{c10,c11}, the two calculations are very 
close, with differences $\Delta \es = 0.8\times 10^{-5}$ and 
 $\Delta M_W = 0.6 \ MeV$.
On the other hand, we see that the Effective approach eliminates the ambiguity associated 
with the choice of the electroweak scale.\\
Inserting in Eqs.(\ref{equ1},\ref{equ2}) the current experimental average
$(\es)_{exp} = 0.23156 \pm 0.00017$ and taking into account the uncertainties
in the other input parameters, we find, for $\dah = 0.02761 \pm 0.00036$, the determination
$M_H = 140^{+97}_{-57}\  GeV$ and the $95 \%$ CL upper bound $M_H^{95}= 334 \ GeV$. On the 
other hand, inserting the current average $(M_W)_{exp} = 80.448 \pm 0.034\  GeV$ in 
Eq.(\ref{equ5}), we obtain $M_H = 25^{+52}_{-25} \ GeV$ and $M_H^{95}= 131 \ GeV$.
The corresponding values for $\dah = 0.02738 \pm 0.00020$, are $M_H =164^{+103}_{-63} \ GeV$, 
$M_H^{95}= 364 \ GeV$ from $\es$, and $M_H =30^{+53}_{-30} \  GeV$, $M_H^{95}= 139 \ GeV$ 
from $M_W$.
We see that the calculation based on the two values of $\dah$ are rather close
and that, at present, the experimental determination of $M_W$ constrains $M_H$
much more sharply than $\es$. In fact, the values of $M_H$ we have obtained 
from $M_W$ are well inside the $95 \%$ CL exclusion region from direct searches 
$M_H < 113 \ GeV$ \cite{c13},
and the corresponding upper bounds leave open but a small range above the exclusion
limit! On the other hand, global analyses are less restrictive and lead, at present, 
to $M_H^{95} = 212 \ GeV$ for $\dah = 0.02761 \pm 0.00036$, and to
$M_H^{95} = 236 \ GeV$ for $\dah = 0.02738 \pm 0.00020$ \cite{c14}.\\
Another interesting application is the use of $\ln{(M_H/100 \ GeV)}$ derived from
$(\es)_{exp}$ and Eqs.(\ref{equ1},\ref{equ2}), to predict $M_W$. In this way we find,
with $\dah = 0.02761 \pm 0.00036$, 
the prediction $M_W = 80.366 \pm 0.026 \ GeV$, 
which differs from $(M_W)_{exp}$ by about $1.9 \sigma$! 
This is the first time over a period of several years that this calculation leads to a 
prediction of $M_W$ that differs from $(M_W)_{exp}$ by more than $1 \sigma$ and, in our
opinion, it is an indication that the current fit to the SM is of lower quality 
than in the past. This is, of course, also strongly indicated by the low confidence level
of the present experimental determination of $\es$.\\
In summary, we have outlined a novel framework to carry out precision calculations in the SM,
in which $\es$ plays the role of the basic electroweak mixing parameter. It shares the 
attractive convergence properties of the $\msbar$ framework and, at the same time, 
eliminates the ambiguities associated with the choice of the electroweak scale.
It may be particularly useful if $\es$ can be measured with an error of $\pm 1 \times 10^{-5}$,
as anticipated in the Tesla project.

The authors are indebted to G.~Degrassi and P.~Gambino for 
detailed discussions and access to their codes. 
This research was supported by NSF grant PHY-$0070787$.

\begin{table}[p] 
\caption{Predicted values of $M_W$ and $\es$ in the 
 EFF renormalization scheme for $M_t=174.3 \ GeV$,
$\hat{\alpha}_s \left( M_Z^2\right) = 0.118$, 
$\Delta \alpha_{had}^{\left( 5 \right)} = 0.02761$, with
QCD corrections based on the $\mu_t$-parametrization. } \label{tab1}
\begin{center}
\begin{tabular}{ |r|| r  | r |} 
\hline 
 & &  \\
$M_H \left[ GeV\right]$ & $M_w \left[ GeV\right]$  &  $\sef $  \\ 
 & & \\
 
\hline \hline
 65  & 80.410 & 0.23117 \\
100  & 80.386 & 0.23139 \\
300  & 80.313 & 0.23196 \\
600  & 80.260 & 0.23235 \\
1000 & 80.220 & 0.23263 \\
\hline
\end{tabular}
\end{center}
\end{table}

\begin{table}[p] 
\caption{As in Table 1, but with $\Delta \alpha_{had}^{\left( 5 \right)}=
0.02738$.} \label{tab2}
\begin{center}
\begin{tabular}{ |r|| r | r |}
\hline 
 & & \\
$M_H \left[ GeV\right]$ &  $M_w \left[ GeV\right]$ &  $\sef $  \\ 
 & &  \\
 
\hline \hline
 65  & 80.414 & 0.23109 \\
100  & 80.391 & 0.23130 \\
300  & 80.317 & 0.23188 \\
600  & 80.264 & 0.23227 \\
1000 & 80.224 & 0.23255 \\
\hline
\end{tabular}
\end{center}
\end{table}

\end{document}